\documentclass{mem}
\usepackage{natbib}\usepackage{txfonts}\usepackage{balance}
\usepackage{graphicx}
\usepackage[a4paper,breaklinks,dvipdfm]{hyperref}
\idline{75}{282}
\begin{document}

\title{Astrometric planet search around M8--L2 dwarfs from the ground and with Gaia}

\author{
J. Sahlmann\inst{1}
		\and P. F. Lazorenko\inst{2}
		\and D. S\'egransan\inst{3}
		\and E. L. Mart\'in\inst{4} 
	        \and M. Mayor\inst{3} 
		\and D. Queloz\inst{3,5} 
		\and S. Udry\inst{3}          }

  \offprints{J. Sahlmann}
 
\institute{European Space Agency, ESAC, P.O. Box 78, Villanueva de la Ca\~nada, 28691 Madrid, Spain. 
		\email{johannes.sahlmann@sciops.esa.int}				
		\and
		Main Astronomical Observatory, Zabolotnogo 27, 03680 Kyiv, Ukraine 
		\and
		Observatoire de Gen\`eve, 51 Chemin Des Maillettes, 1290 Versoix, Switzerland 
		\and  
		INTA-CSIC Centro de Astrobiolog\'ia, 28850 Torrej\'on de Ardoz, Madrid, Spain
		\and
		University of Cambridge, Cavendish Laboratory, J J Thomson Avenue, Cambridge, UK}

\authorrunning{Sahlmann et al.}
\titlerunning{Astrometric planet search around UCDs}

\abstract{Ultra-cool dwarfs are very low-mass stars or brown dwarfs and because of their faintness they are difficult targets for radial velocity and transit planet searches. High-precision astrometry is one way to efficiently discover planets around these objects. We are conducting a planet search survey of 20 M8--L2 using ground-based imaging astrometry with FORS2 at VLT. The realised accuracy of 100 micro-arcseconds allows us to set stringent constraints on the presence of planets, to discover astrometric binaries, and to measure parallaxes with an unprecedented precision of 0.1 \%. The obtained detection limits firmly establish that giant planets are rare around UCDs at all separations. The astrometric performance of our programme is comparable to what is expected from \emph{Gaia} observations of single faint objects and we discuss potential synergies for planet searches around ultracool dwarfs. We estimate that \emph{Gaia} will be able to characterise $\sim$100 astrometric binaries with an ultracool primary.

\keywords{Stars: low-mass -- Brown dwarfs -- Planetary systems -- Binaries: close  -- Astrometry -- Parallaxes} }
\maketitle{}

\section{Introduction}
Very low-mass stars and brown dwarfs are becoming popular targets for exoplanet searches mainly for two reasons: (1) They allow us to study the planet formation process around the lowest-mass products of star formation and (2) they may provide for a environment in which 'habitable' exoplanets can reside. Because of their faintness and physical properties, these ultracool dwarfs (UCD) make it difficult to obtain observations that can efficiently detect exoplanets around them.

Since 2010, we are pursuing a planet search around 20 ultracool dwarfs of M8--L2 spectral types using the astrometry technique. An overview of the project and first results that include the parallaxes and planet detection limits are reported in \cite{Sahlmann:2014aa}. The data reduction techniques and a deep astrometric catalogue are presented in \cite{Lazorenko:2014aa}.

The \emph{Gaia} satellite was launched in December 2013 and begins its all-sky survey around mid-2014. Typical UCDs lie at the faint end of the \emph{Gaia} magnitude range, but up to a few thousand of them will be observed. 

\section{First results of our ground-based programme}
After two years of survey duration, we have completed the initial screening of the target sample and could derive first constraints on the occurrence of planets around UCD.

\subsection{Giant planet occurrence}
The combination of detection technique (astrometry), single-measurement uncertainty ($\sim$120 micro-arcseconds or $\mu$as), and sampling (10 epochs over a 16 months time-span, on average) means that we are most sensitive to giant planets in intermediate-period orbits. So far, no planetary-mass companion was detected, but we are following up on promising candidates by obtaining additional observations. Using the uniform data of the first two survey years, we found that less than 9 \% of M8--L2 dwarfs host giant planets with masses $\gtrsim$5$M_J$ and separations of 0.1--0.8 AU \citep{Sahlmann:2014aa}. 

With this result of our study, it is now firmly established that giant planets are rare around UCDs at all separations.

\subsection{Parallaxes and distances}
Astrometric planet search relies on detecting periodic deviations from the standard astrometric motion. Therefore, the determination of the five standard parameters - positions, parallax, proper motions - is required in the process. We thus measured the relative parallaxes of the 20 targets and absolute parallaxes were obtained with an additional correction step. For most targets, this represented the first parallax determination. The excellent astrometric precision translates into parallax uncertainties of 0.1 mas and smaller, which is unprecedented for optical observations of UCDs. Consequently, the absolute magnitude uncertainty for these 20 M/L dwarfs is now dominated by the photometric measurement and no longer by the distance determination.

\subsection{New tight binaries}
\begin{figure}[t!]
\resizebox{\hsize}{!}{\includegraphics[clip=true]{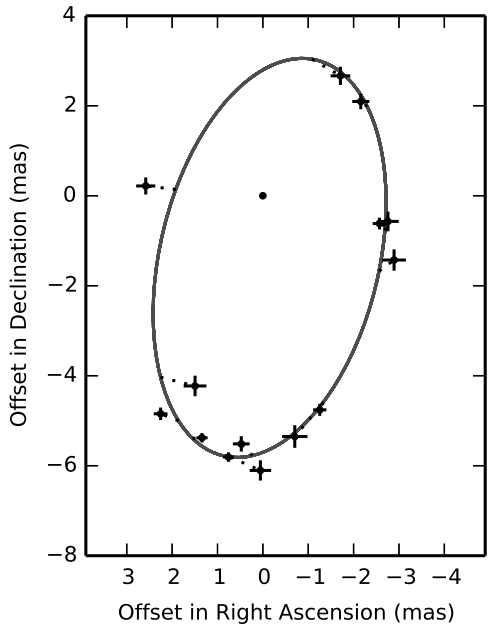}}
\caption{\footnotesize
The astrometric orbit of DE0823--49 caused by its low-mass companion \citep{Sahlmann:2013ab}. The orbital period is 246 days and the orbital signature is $\sim$ 4 mas. The estimated magnitude of the primary is $G\!\sim$18.7, therefore this orbit will also be detected with \emph{Gaia} astrometry.
}
\label{fig:orbit}
\end{figure}

In two cases, we detected periodic residuals after adjusting the standard astrometric model. These deviations have periods of several hundred days and amplitudes of several milli-arcseconds (mas), thus must be caused by a massive orbiting companion, see Fig. \ref{fig:orbit}. The companion of DE0823--49 was found to have a mass of $28\pm2 \,M_J$ (for an age of 1 Gyr) and this system is one of the very few known that have a small mass-ratio and a separation well below 1 AU \citep{Sahlmann:2013ab}. The second tight binary discovered in our survey (DE0630--18) will soon be fully characterised.

\subsection{Deep astrometric catalogue}
\begin{figure}[t!]
\resizebox{\hsize}{!}{\includegraphics[clip=true]{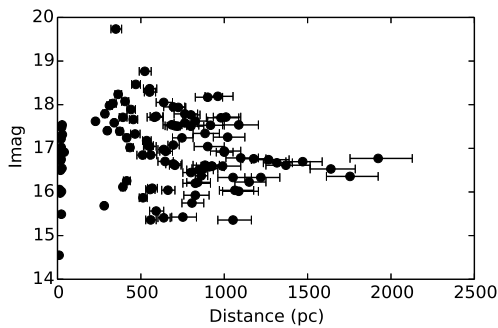}}
\caption{\footnotesize
Magnitudes and distances of a subset of 113 objects in the astrometric catalogue of \citet{Lazorenko:2014aa} that have $\chi^2<2$ and for which the parallax is known at better than 10 \%. The UCDs are located at the left margin of the figure and the most remote reference star is at a distance of 2 kpc.
}
\label{fig:palta2}
\end{figure}

As detailed by \cite{Lazorenko:2014aa}, the astrometric reduction technique relies on the availability of many reference stars. During the reduction process, the astrometric parameters of these stars are determined as well. We made the catalogue of reference stars publically available (VizieR J/A+A/565/A21). It covers the 20 target fields of roughly $2\arcmin \times 2\arcmin$ size and a magnitude range of $I\sim14-22$. Figure \ref{fig:palta2} shows a subset of reference stars with well-determined distances, which could serve as a comparison sample for \emph{Gaia} astrometry of faint stars.

\section{Astrometric orbits of ultracool dwarfs with Gaia}
The expected yield of \emph{Gaia} observations of ultracool dwarfs is discussed by \cite{de-Bruijne14} and \cite{Smart14} in this volume (see also \citealt{Sarro:2013uq}). Accurate distances of some thousand UCDs will be obtained from \emph{Gaia}'s absolute parallax measurements, where UCDs that are companions to stars are not considered. For some of them, \emph{Gaia} will detect orbital motion of the photocentre.

\subsection{Binaries with UCD primaries}
About 10--30 \% of UCDs are found in multiple systems \citep{Burgasser:2007ix}. The orbits of binaries with separations $\gtrsim$3 AU are unlikely to be characterised by  \emph{Gaia}, because their orbital periods exceed twice the nominal mission lifetime of 5 years. In the survey of nearby 20 M/L dwarfs, we found that $10^{+11}_{-3}$ \% of UCD binaries are tight enough ($\lesssim1$ AU) and have photocentric semi-amplitudes large enough ($\gtrsim1$ milli-arcsec) to be detectable by \emph{Gaia} \citep{Sahlmann:2014aa}. \cite{Smart14} estimates that \emph{Gaia} with the nominal $G<20$ limit will observe about 500 L and T dwarfs, the majority being early-L dwarfs. Since almost all of these are also nearby $<50$ pc, we can expect that 30--105 orbits of UCD binaries with an L/T primary will be detected and possibly characterised by \emph{Gaia}. In addition, several hundred nearby late-M dwarfs will be observed, of which 6--21 \% can also be expected to be detectable astrometric binaries.

In summary, \emph{Gaia} will detect the astrometric orbital motion of about 100 -- 200 UCDs with spectral types later than M8. For many of them, it will be possible to determine the orbital parameters from \emph{Gaia} astrometry alone, which will result in an important sample to study the properties of very low-mass binary systems. However, many will rely on follow-up observations to obtain a detailed characterisation, e.g.\ to refine the orbital properties or to relate photocentric and barycentric orbit sizes to constrain the individual component masses.
 
\subsection{Planets around UCDs}
It is unlikely that \emph{Gaia} will detect many planets around UCDs. Whereas the astrometric signature of a Jupiter-mass planet in a 1000 day orbit around WISE J1049--5319A (L8 at 2 pc, \citealt{Luhman:2013aa}) is $\sim$8~mas and readily detectable, it drops to $\sim$0.5~mas for a typical UCD (L2 at 20 pc). Because of the targets' faintness, the \emph{Gaia} single-measurement precision will be of the same order of magnitude and consequently it will be difficult to characterise the planet's signature with the typical number of $\sim$70 measurements per object over the mission lifetime. However, it will be possible to identify UCDs with excess astrometric residuals that hint towards the presence of a planetary-mass companion.

These candidates will be ideal targets for follow-up astrometric observations with large ground-based optical telescopes. With the astrometric performance demonstrated with FORS2/VLT, it will be possible to detect and characterise the orbital motion, if present. The detection of extrasolar planets around ultracool dwarfs is thus another example of a synergy between \emph{Gaia} and ground-based observations.

\section{Conclusions}
Using observations with FORS2/VLT, we have shown how 100 $\mu$as astrometry can be used to uniquely constrain the presence of planets around nearby ultracool dwarfs. In terms of distance determination and binary detection, our ground-based programme demonstrates what can be expected from the \emph{Gaia} observations of UCDs. Extrapolating our findings of M8-L2 dwarfs, we predict that \emph{Gaia} could characterise the orbits of about 100 binary systems having an ultracool primary. Planet detections with \emph{Gaia} are possible, but will be limited to very few special cases. On the other hand, \emph{Gaia} will provide us with a list of planet candidates around UCDs that will have to be confirmed with complementary observations.

\begin{acknowledgements}
J.S. is supported by an ESA research fellowship.
\end{acknowledgements}

\bibliographystyle{aa}

\begin{thebibliography}{8}
\expandafter\ifx\csname natexlab\endcsname\relax\def\natexlab#1{#1}\fi

\bibitem[{{Burgasser} {et~al.}(2007){Burgasser}, {Reid}, {Siegler}, {Close},
  {Allen}, {Lowrance}, \& {Gizis}}]{Burgasser:2007ix}
{Burgasser}, A.~J., {Reid}, I.~N., {Siegler}, N., {et~al.} 2007, Protostars and
  Planets V, 427

\bibitem[{{de Bruijne}(2014)}]{de-Bruijne14} {de Bruijne}, J.~H.~J. 2014, \memsai, 85, xx

\bibitem[{{Lazorenko} {et~al.}(2014){Lazorenko}, {Sahlmann}, {S{\'e}gransan},
  {Mart{\'{\i}}n}, {Mayor}, {Queloz}, \& {Udry}}]{Lazorenko:2014aa}
{Lazorenko}, P.~F., {Sahlmann}, J., {S{\'e}gransan}, D., {et~al.} 2014, \aap,
  565, A21

\bibitem[{{Luhman}(2013)}]{Luhman:2013aa}
{Luhman}, K.~L. 2013, \apjl, 767, L1

\bibitem[{{Sahlmann} {et~al.}(2014){Sahlmann}, {Lazorenko}, {S{\'e}gransan},
  {Mart{\'{\i}}n}, {Mayor}, {Queloz}, \& {Udry}}]{Sahlmann:2014aa}
{Sahlmann}, J., {Lazorenko}, P.~F., {S{\'e}gransan}, D., {et~al.} 2014, \aap,
  565, A20

\bibitem[{{Sahlmann} {et~al.}(2013){Sahlmann}, {Lazorenko}, {S{\'e}gransan},
  {Mart{\'{\i}}n}, {Queloz}, {Mayor}, \& {Udry}}]{Sahlmann:2013ab}
{Sahlmann}, J., {Lazorenko}, P.~F., {S{\'e}gransan}, D., {et~al.} 2013, \aap,
  556, A133

\bibitem[{{Sarro} {et~al.}(2013){Sarro}, {Berihuete}, {Carri{\'o}n}, {Barrado},
  {Cruz}, \& {Isasi}}]{Sarro:2013uq}
{Sarro}, L.~M., {Berihuete}, A., {Carri{\'o}n}, C., {et~al.} 2013, \aap, 550,
  A44

\bibitem[{{Smart}(2014)}]{Smart14} {Smart}, R.~L. 2014, \memsai, 85, xx

\end{thebibliography}

\end{document}